\documentclass[12pt]{article}
\pdfoutput=1

\usepackage{amsmath}
\usepackage{amssymb}
\usepackage{epsfig}
\usepackage{epstopdf}
\usepackage{latexsym}
\numberwithin{equation}{section}
\newcommand{\be}{\begin{equation}}
\newcommand{\ee}{\end{equation}}
\newcommand{\bea}{\begin{eqnarray}}
\newcommand{\eea}{\end{eqnarray}}

\newcommand{\nn}{\nonumber}

\usepackage[
      colorlinks=false,
      linkcolor=darkblue,  
      urlcolor=blue,    
      filecolor=blue,     
      citecolor=red,
linktocpage=true,
      pdfstartview=FitV,
      bookmarksopen=true    
      ]{hyperref}

\begin{document}

\begin{titlepage}

\centerline{\Huge \rm Supersymmetric AdS$_6$ solutions} 
\bigskip
\centerline{\Huge \rm of type IIB supergravity}
\bigskip
\bigskip
\bigskip
\bigskip
\bigskip
\bigskip
\centerline{\rm Hyojoong Kim$^a$, Nakwoo Kim$^a$ and Minwoo Suh$^b$}
\bigskip
\centerline{\it $^a$Department of Physics and Research Institute of Basic Science}
\centerline{\it Kyung Hee University, Seoul 130-701, Korea} 
\bigskip 
\centerline{\it $^b$Department of Physics}
\centerline{\it Sogang University, Seoul 121-742, Korea}  
\bigskip
\centerline{\tt h.kim@khu.ac.kr, nkim@khu.ac.kr, minsuh@usc.edu} 
\bigskip
\bigskip
\bigskip
\bigskip
\bigskip
\bigskip
\bigskip
\bigskip
\bigskip
\bigskip

\begin{abstract}
We study the general requirement for supersymmetric AdS$_6$ solutions in type IIB supergravity. We employ the Killing spinor technique and study the differential and algebraic relations among various Killing spinor bilinears to find the canonical form of the solutions. Our result agrees precisely with the work of Apruzzi et. al. \cite{Apruzzi:2014qva} which used the pure spinor technique. We also obtained the four-dimensional theory through the dimensional reduction of type IIB supergravity on AdS$_6$. This effective action is essentially a nonlinear sigma model with five scalar fields parametrizing $\textrm{SL}(3,\mathbb{R})/\textrm{SO}(2,1)$, modified by a scalar potential and coupled to Einstein gravity in Euclidean signature. We argue that the scalar potential can be explained by a subgroup CSO(1,1,1) $\subset\textrm{SL}(3,\mathbb{R})$ in a way analogous to gauged supergravity. 
\end{abstract}

\vskip 4cm

\flushleft {June, 2015}

\end{titlepage}

\tableofcontents

\vspace{3cm}

\section{Introduction}

In recent years, there have been renewed interests in supersymmetric AdS$_6$ solutions in $D=10$ supergravity. Via the gauge/gravity correspondence \cite{Maldacena:1997re}, such solutions should be dual to certain $D=5$ superconformal field theories. Five-dimensional gauge theories are perturbatively non-renormalizable. Seiberg nonetheless argued that $\mathcal{N}$ = 1 supersymmetric $Sp(N)$ gauge theories with hypermultiplets of $N_f<$8 fundamental and one antisymmetric tensor representation flow in the infinite gauge coupling limit to superconformal theories, and their $SO(N_f){\times}U(1)$ global symmetry is enhanced to $E_{N_f+1}$ \cite{Seiberg:1996bd, Morrison:1996xf, Intriligator:1997pq}. Such fixed point theories have string theory construction: in terms of the near-horizon limit of D4-D8 brane configurations. Based on the AdS$_6$/CFT$_5$ correspondence \cite{Ferrara:1998gv}, Brandhuber and Oz identified the gravity dual as supersymmetric AdS$_6 \times_w S^4$ solution of massive type IIA supergravity \cite{Brandhuber:1999np}. More recently this correspondence was generalized to quiver gauge theories and AdS$_6 \times_w S^4/\mathbb{Z}_n$ orbifolds in \cite{Bergman:2012kr}.

Thanks to the development of the localization technique \cite{Pestun:2007rz} and its generalization to five-dimensional gauge theories \cite{Hosomichi:2012ek, Kallen:2012va}, some BPS quantities can be calculated exactly. The conjectured enhancement of global symmetry to  $E_{N_f+1}$ was verified from the analysis of superconformal index in \cite{Kim:2012gu}. Furthermore, the $S^5$ free energy  and also the $\frac{1}{2}$-BPS circular Wilson loop operators are calculated and shown to agree with the gravity side computations \cite{Jafferis:2012iv,Assel:2012nf,Alday:2014rxa,Alday:2014bta}. 

Encouraged by the successful application of localization technique on the field theory side, it is natural for us to look for new supersymmetric AdS$_6$ solutions. In massive type IIA supergravity, it was proved that the Brandhuber-Oz solution is the unique one \cite{Passias:2012vp}.  In type IIB supergravity, the T-dual version of the Brandhuber-Oz solution has been known for a long time \cite{Cvetic:2000cj}. A new solution was obtained more recently employing the technique of  non-Abelian T-dual transformation in \cite{Lozano:2012au}. The dual gauge theory was investigated in \cite{Lozano:2013oma}, but it is not completely understood yet. 

For a thorough study, the authors of \cite{Apruzzi:2014qva} investigated the general form of supersymmetric AdS$_6$ solutions of type IIB supergravity, using \emph{the pure spinor} approach. They found that the four-dimensional internal space is a fibration of $S^2$ over a two-dimensional space, and also showed that the supersymmetry conditions boil down to two coupled partial differential equations. Of course any solution of the PDEs provides a supersymmetric AdS$_6$ solution at least locally. In particular, the two explicit solutions mentioned above can be reproduced as specific solutions to the PDEs. But otherwise these non-linear coupled PDEs are so complicated that currently it looks very hard, if not impossible, to obtain more AdS$_6$ solutions by directly solving the PDEs.

The objective of this article is to procure additional insight into this problem, using alternative methods. In the first part we use the \emph{Killing spinor} approach which is probably more well-known and has been successfully applied to many similar problems, see $e.g.$ \cite{Gauntlett:2004zh, Lin:2004nb, Gauntlett:2005ww}. Following the standard procedure we work out the algebraic and differential constraints which should be satisfied by various spinor bilinears and derive the supersymmetric conditions.  In the end, we confirm that our results are in precise agreement with that of \cite{Apruzzi:2014qva}. Secondly, via dimensional reduction of the bosonic sector of the $D=10$ action on AdS$_6$, we present a four-dimensional effective theory action, which turns out to be a non-linear sigma model of five scalar fields coupled to gravity. The scalar fields parametrize the coset space $\textrm{SL}(3,\mathbb{R})/\textrm{SO}(2,1)$. Also there is a non-trivial scalar potential, which breaks the global $sl(3,\mathbb{R})$ symmetry to a certain subalgebra. Although in this paper we do not present new solutions, we believe the identification of the $D=4$ effective action will prove useful in the construction of explicit solutions and their classifications. 

This paper is organized as follows. Section \ref{susy} contains an analysis on the supersymmetry conditions for AdS$_6$ solutions. In section \ref{nlsm}, we study the four-dimensional effective theory from dimensional reduction on AdS$_6$. In section 4 we conclude. Technical details are relegated to appendices.

\section{Supersymmetric  AdS$_6$ solutions} \label{susy}

\subsection{Killing spinor equations}

We consider the most general supersymmetric AdS$_6$ solutions of type IIB supergravity. We take the D=10 metric as a warped product of AdS$_6$ with a four-dimensional Riemannian space $M_4$
\be
ds^2= e^{2U} ds_{AdS_6}^2+ ds_{M_4}^2,
\ee
where $U$ is a warp factor. To respect the symmetry of AdS$_6$, we should set the five-form flux to zero. The complex three-form flux $G$ is non-vanishing only on $M_4$. The warp factor $U$, the dilation $\phi$ and the axion $C$, are functions on $M_4$ and independent of coordinates in AdS$_6$.

To preserve some supersymmetry, we require the vanishing of supersymmetry transformations of the gravitino and the dilatino $i.e.$ $\delta \psi_M=0,~ \delta \lambda=0$. With the gamma matrix decomposition \eqref{gamma} and the spinor ansatz \eqref{spinor}, we reduce the ten-dimensional Killing spinor equations to four-dimensional ones. There are two differential and four algebraic-type equations:
\begin{align}
D_m\xi_{1\pm}+\frac{1}{96}G_{npq}(\gamma_m \gamma^{npq}+2\gamma^{npq}\gamma_m)\xi_{2\pm}&=0,\label{dif-gra1} \ \\
\bar{D}_m\xi_{2\pm}+\frac{1}{96}G^*_{npq}(\gamma_m \gamma^{npq}+2\gamma^{npq}\gamma_m)\xi_{1\pm}&=0,\label{dif-gra2} \\
im e^{-U}\xi_{1\mp}+\partial_n U \gamma^n\xi_{1\pm}-\frac{1}{48}G_{npq}\gamma^{npq}\xi_{2\pm}&=0,\label{alg-gra1} \\
im e^{-U}\xi_{2\mp}+\partial_n U\gamma^n\xi_{2\pm}-\frac{1}{48}G^*_{npq}\gamma^{npq}\xi_{1\pm}&=0, \label{alg-gra2}\\
P_n\gamma^n\xi_{2\pm}+\frac{1}{24}G_{npq}\gamma^{npq}\xi_{1\pm}&=0, \label{alg-dil1}\\
P^*_n\gamma^n\xi_{1\pm}+\frac{1}{24}G^*_{npq}\gamma^{npq}\xi_{2\pm}&=0,\label{alg-dil2}
\end{align}
where
\be
D_m\xi_{1\pm} = (\nabla_m -\dfrac{i}{2} Q_m )\xi_{1\pm}, \qquad \bar{D}_m\xi_{2\pm} = (\nabla_m +\dfrac{i}{2} Q_m )\xi_{2\pm}.
\ee
With the assumption that there exists at least one nowhere-vanishing solution to the equations in the above, we can construct various spinor bilinears. Then the supersymmetric condition is translated into various algebraic and differential relations between the spinor bilinears. We have recorded them in appendix \ref{alg-rel} and \ref{diff-rel}.

\subsection{Killing vectors}

We first need to study the isometry of the four-dimensional Riemannian space $M_4$. We note that the following two complex vectors satisfy the Killing equation $\nabla_{(m} K_{n)}=0$.
\begin{align}
\overline{\xi}_{1+}\gamma_{n}\,\xi_{1-}+\, \overline{\xi}_{2+}\gamma_{n}\,\xi_{2-}, \qquad
\overline{\xi^c_1}_{+}\gamma_{n}\,\xi_{2-}+\,\overline{\xi^c_2}_{+}\gamma_{n}\,\xi_{1-}.
\end{align}
If these vectors are to provide a true symmetry of the full ten-dimensional solution as well, we need to check if
\be
\mathcal{L}_K \,U = (d \,i_K +i_K \,d) \,U=K^m \partial_m U=0,
\ee
where $\mathcal{L}_K $ is a Lie derivative along the Killing vector $K$. From \eqref{du.vec} and \eqref{du.cvec}, we find that in fact only three of them satisfy the above condition. Hence, the true Killing vectors are
\begin{align}
K_1^{n} & \equiv \textrm{Re} \,(\overline{\xi^c_1}_{+}\gamma^{n}\,\xi_{2-}+\,\overline{\xi^c_2}_{+}\gamma^{n}\,\xi_{1-}),\\
K_2^n & \equiv \textrm{Im}\, (\overline{\xi^c_1}_{+}\gamma^{n}\,\xi_{2-}+\,\overline{\xi^c_2}_{+}\gamma^{n}\,\xi_{1-}),\\
K_3^n & \equiv \textrm{Re}\, (\overline{\xi}_{1+}\gamma^{n}\,\xi_{1-}+\, \overline{\xi}_{2+}\gamma^{n}\,\xi_{2-}).
\end{align}
Using \eqref{alg-dil1} and \eqref{alg-dil2}, we have $P_m \,K_i^m=0$, which implies that 
\be
\mathcal{L}_{K_i} \phi = \mathcal{L}_{K_i} C=0,
\ee
where $i=1,2,3$. Also we obtain $i_K *G=0$ from \eqref{BPS4} and \eqref{BPS5}, and $i_k d*G=0$ using the equation of the motion for G,\footnote{The equation of the motion for $G$ is $d*G= (-6 dU + i Q)\wedge *G +P \wedge *G^*$. \label{eomG}} thus 
\be
\mathcal{L}_{K_i} * G=0.
\ee
Hence, we conclude that $K_i$ describe symmetries of the full ten-dimensional solutions. 

Now let us study the Lie bracket of the Killing vectors. Using \eqref{dK3} and \eqref{d(K1+iK2)}, the Fierz identities \eqref{fierz1.2} and the normalization \eqref{sc-norm}, we show that the three Killing vectors satisfy an $SU(2)$ algebra,
\be
[K_i\, , K_j]= \epsilon_{ijk} K_k.
\ee
This $SU(2)$ isometry of the four-dimensional Riemannian space corresponds to the $SU(2)_R$ R-symmetry of dual five-dimensional field theory. Then we  construct a $3 \times 3$ matrix, whose elements are the inner products of the Killing vectors \eqref{innerK}, and find that this matrix is singular
\be
\textrm{det} \, (K_i \cdot K_j)=0.
\ee
This guarantees that $K_i$ are the Killing vectors of $S^2$. The radius $l$ of the two-sphere is given by
\be
2 l^2=(K_1)^2+(K_2)^2+(K_3)^2= 2\left[\frac{1}{9m^2}e^{2U}-4(\overline{\xi}_{1+}\xi_{2+})(\overline{\xi}_{2+}\xi_{1+})\right].
\ee

\subsection{Supersymmetric solutions}

We have showed that once we require the supersymmetry conditions, then the four-dimensional Riemannian space should contain $S^2$. Now we focus on the remaining two-dimensional space. We start with two one-forms $L^1_n$ and $L^2_n$ from \eqref{S21}
\begin{align}
L^1_n\,&\equiv\, e^{U+\frac{1}{2}\phi}(\overline{\xi}_{1+}\xi_{2+}+\overline{\xi}_{2+}\xi_{1+})\,\partial_n C -m e^{-\frac{1}{2}\phi} \,L^3_n, \nn\\
&= -i \partial_n \Big(e^{U-\frac{1}{2}\phi}(\overline{\xi}_{1+}\xi_{2+}-\overline{\xi}_{2+}\xi_{1+})\Big),\\
L^2_n\,&\equiv\,\textrm{Im}\left(\overline{\xi}_{1+}\gamma_n\xi_{2-}\,+\,\overline{\xi}_{2+}\gamma_n\xi_{1-}\right)= \dfrac{1}{m} e^{-\frac{1}{2}\phi} \partial_n \Big(e^{U+\frac{1}{2}\phi}(\overline{\xi}_{1+}\xi_{2+}+\overline{\xi}_{2+}\xi_{1+})\Big),
\end{align}
where
\be\label{L3}
L^3_n = \textrm{Re}\left(\overline{\xi}_{1+}\gamma_n\xi_{2-}\,-\,\overline{\xi}_{2+}\gamma_n\xi_{1-}\right).
\ee
Using the Fierz identities, one can show that the one-forms $L^2$ and $L^3$ are orthogonal to the Killing vectors
\be
K_i \cdot L^2=K_i \cdot L^3=0.
\ee
Together with ${\cal{L}}_{K_i}C=0$, the one-form $L^1$ is also orthogonal to the Killing vectors. Then, we introduce coordinates $z$ and $y$,
\begin{align}\label{coord}
z&=-3m i \,e^{U-\frac{1}{2}\phi}(\overline{\xi}_{1+}\xi_{2+}-\overline{\xi}_{2+}\xi_{1+}),\\
y&=3m\, e^{U+\frac{1}{2}\phi}(\overline{\xi}_{1+}\xi_{2+}+\overline{\xi}_{2+}\xi_{1+}). \nn
\end{align}
Since ${\cal{L}}_{K_i}z=i_{K_i} dz\sim K_i \cdot L^1=0$ and similarly ${\cal{L}}_{K_i}y=0$, 
the coordinates $z$ and $y$ are independent of the sphere coordinates. In terms of the coordinates $z$ and $y$, the one-forms are
\begin{align}
L^1 &=\frac{1}{3m}y dC- m e^{-\frac{1}{2}\phi} L^3=\frac{1}{3m}dz,\label{L1}\\
L^2 &=\frac{1}{3m^2}e^{-\frac{1}{2}\phi}dy\label{L2}.
\end{align}

Then we calculate inner products of the one-forms $L^1$ and $L^2$, hoping to be able to fix the remaining two-dimensional metric. However, we cannot immediately calculate the inner products involving $L^1$, because it includes $dC$. The resolution is that we consider the one-form $L^3$ defined in \eqref{L3} instead. From \eqref{d(iL2-L3)} and \eqref{d(iL2+L3)}, we have 
\begin{align}
d(e^{4U-\frac{1}{2}\phi} L^2) &=e^{4U+\frac{1}{2}\phi} dC \wedge L^3, \label{dL2}\\
d(e^{4U+\frac{1}{2}\phi}\,L^3)&=0\label{dL3}.
\end{align}
We introduce another coordinate $w$ and write $L^3$ as
\be
L^3= \frac{1}{3m^2}e^{-4U-\frac{1}{2}\phi}\, dw.
\ee
Then we can calculate inner products of $L^2$ and $L^3$ using the Fierz identities and read off the two-dimensional metric components in $w$ and $y$ coordinates,
\begin{align}
ds^2_2=\frac{1}{m^2(e^{4U+\phi}-y^2-e^{2\phi}z^2)}\Big[&e^{-2U+\phi}(e^{4U-\phi}-z^2)dy^2 \\
            &+ e^{-10U-\phi}(e^{4U+\phi}-y^2)dw^2  -2e^{-6U} \,y \,z \,dy\, dw \Big]. \nn
\end{align}
At this stage, $z$ is an unknown function of $y$ and $w$. The details are in appendix \ref{innervec}.

We would like to express $dC$ in terms of the coordinate $z$ instead of $w$.
From the Killing spinor equations 
\eqref{alg-gra1}--\eqref{alg-dil2},
 we have
\bea
L_2 \cdot dC &=& e^{-\phi}d(4U+\phi)\cdot L_3-\frac{4}{3}e^{-2U-\frac{1}{2}\phi}z, \label{L2dC}\\
L_3\cdot dC &=& e^{-\phi}d(4U-\phi)\cdot L_2+\frac{4}{3}e^{-2U-\frac{3}{2}\phi}y. \label{L3dC}
\eea
The integrability conditions $d(dz)= d(dy)=0$ from \eqref{L1}, \eqref{L2}, when combined with \eqref{dL2}, \eqref{dL3} give
\begin{align}
L_2 \wedge dC+e^{-\phi}d(4U+\phi) \wedge L_3=0,\label{consistency1}\\
L_3 \wedge dC+e^{-\phi}d(4U-\phi) \wedge L_2=0.\label{consistency2}
\end{align}
Summarising, from \eqref{L2dC}--\eqref{consistency2}, we find that 
\be
dC=\frac{1}{2 y z} \Big[(e^{4U-\phi}-e^{-2\phi}y^2)\,d(4U-\phi)+(e^{4U-\phi}-z^2)\,d(4U+\phi)
                                          -4 e^{-4U-\phi}z\,dw +4 e^{-2\phi} y\, dy\Big].
\ee
If we plug this into \eqref{L1}, we can express $dw$ in terms of $dy$ and $dz$. Then, we can write the metric and $dC$ in the $y$ and $z$ coordinates.

Now we are ready to present our main result. We introduce a new coordinate $x$ defined by
\be
x^2=e^{8U}-e^{4U-\phi}y^2-e^{4U+\phi}z^2.
\ee
Then, we can have all fields and functions in terms of coordinates $x$ and $y$ only. We have the metric of the four-dimensional Riemannian space,
\begin{align} \label{metric0}
ds^2_4\,=&\,\frac{1}{9m^2}\Big[\, e^{-6U}\,x^2\,ds^2_{S^2}\,   \\
&+\,\frac{e^{-2U}}{e^{8U+\phi}-e^{\phi}x^2-e^{4U}y^2}\,
\Big[(e^{4U+\phi}-y^2)\,dx^2+9\,(e^{8U}-x^2)\,dy^2+6\,x\,y\,dx\,dy\Big]\Big]\,. \nn
\end{align}
Similarly $dC$ is written as
\begin{align} \label{BPS3-axion}
dC=\frac{e^{-2U-\phi}}{y\sqrt{e^{8U+\phi}-e^{\phi}x^2-e^{4U}y^2}}\Big[&2(e^{8U+\phi}+e^{\phi}x^2)dU \\
&-\frac{1}{2}(e^{8U+\phi}-e^{\phi}x^2-2e^{4U}y^2)d\phi-\frac{2}{3}e^{\phi}xdx\Big]\,.\nn
\end{align}
The consistency conditions \eqref{consistency1} and \eqref{consistency2} give two partial differential equations,
\begin{align}
4 e^\phi x= 12\Big(e^{8U+\phi}+e^\phi x^2-2 e^{4U} y^2\Big)\, &\partial_x U+8 e^\phi x y \,\partial_y U  \nn \\
-3e^\phi\Big(e^{8U}-x^2\Big)\,&\partial_x\phi+ 2 e^\phi x y \,\partial_y \phi, \label{BPS1}\\
-4  e^{4U+\phi} x y=12e^{4U} y\Big(e^{8U+\phi}-3e^\phi x^2-2 e^{4U}y^2\Big)\,&\partial_x U+4 e^{2\phi} x\Big(e^{8U}+x^2\Big) \,\partial_y U \nn \\
+e^{\phi} x\Big(-e^{8U+\phi}+e^\phi x^2+2 e^{4U}y^2\Big) \,&\partial_y \phi- 3 y e^{4U+\phi}\Big(e^{8U}-x^2\Big)\, \partial_x \phi. \label{BPS2}
\end{align}
The complex three-form flux is obtained by using \eqref{alg-gra1}--\eqref{alg-dil2} rather straightforwardly,
\begin{align}
*\textrm{Re}\,G&=-\frac{2}{y}e^{-6U-\phi/2}  \label{BPS4} \\ 
&\times\left[(e^{8U+\phi}+e^{\phi}x^2+2e^{4U}y^2)dU-\frac{1}{4}(e^{8U+\phi}-e^{\phi}x^2)d\phi-\frac{1}{3}e^{\phi}xdx-2e^{4U}ydy\right]\,,  \nn\\ 
*\textrm{Im}\,G&=2\frac{e^{-4U-\phi/2}}{\sqrt{e^{8U+\phi}-e^{\phi}x^2-e^{4U}y^2}}\label{BPS5}\\
 &\times\left[(3e^{8U+\phi}-e^{\phi}x^2-2e^{4U}y^2)dU+\frac{1}{4}(e^{8U+\phi}-e^{\phi}x^2)d\phi+\frac{1}{3}e^{\phi}xdx+2e^{4U}ydy\right]\,. \nn
\end{align}
Here we used $\gamma_{mnpq}=\sqrt{g_4}\epsilon_{mnpq}\gamma_5$.
 
To summarize, we have employed \emph{the Killing spinor analysis} in Einstein frame and obtained the most general supersymmetric AdS$_6$ solutions for the metric and the fluxes in terms of the warping factor $U$ and the dilation $\phi$. This implies that, when we have solutions $U$ and $\phi$ to the two PDEs \eqref{BPS1} and \eqref{BPS2}, then we can completely determine the metric \eqref{metric0}, the one-form flux \eqref{BPS3-axion} and the three-form flux \eqref{BPS4}, \eqref{BPS5}. Our analysis shows a perfect agreement with the work of \cite{Apruzzi:2014qva}, where the authors used \emph{the pure spinor approach} in string frame. We can reproduce their results with the following identification of our fields to theirs.
\be
g_{mn} \rightarrow e^{-\frac{\phi}{2}} g_{mn}, \quad
U\rightarrow A-\frac{\phi}{4}, \quad 
dC \rightarrow F_1, \quad 
\textrm{Re}\, G \rightarrow e^{-\frac{\phi}{2}} H_3, \quad 
\textrm{Im}\, G \rightarrow -e^{\frac{\phi}{2}} F_3.
\ee
Also our coordinates $(x, y)$ correspond to $(p, q)$ defined in (4.17) of \cite{Apruzzi:2014qva}.

\subsection{Equations of motion}

From the equations of motion and the Bianchi identities of D=10 type IIB supergravity, we obtain the four-dimensional ones via dimensional reduction. Let us start with dualizing the complex three-form flux $G$ into real scalars $f$ and $g$
\begin{align}
*\, \textrm{Re} \,G&= \frac{1}{2} e^{-6U+\frac{1}{2}\phi}(Cdf-fdC+d\tilde{g}),\nn \\
                              &= \frac{1}{2} e^{-6U+\frac{1}{2}\phi}(dg+2C df),\\
*\, \textrm{Im} \,G&= e^{-6U-\frac{1}{2}\phi} df.
\end{align}
where $g=\tilde{g}-fC$. They satisfy the equation of motion for $G$ automatically. Also the Bianchi identity for $P$ is satisfied by \eqref{PQ}. Then the Einstein equation, the equation for $P$ and the Bianchi identity for $G$ give the following six equations.
\begin{align}\label{ein-eom}
R_{mn} = 6 \nabla_m \nabla_n U + 6 \partial_m U \partial_n U +\frac{1}{2}e^{2\phi} \partial_m C \partial_n C +\frac{1}{2} \partial_m \phi \partial_n \phi \nn \\
-\frac{1}{8}\Big[e^{-12U+\phi} \Big((\partial_m g+ 2C \partial_m f)(\partial_n g+ 2C \partial_n f)-\frac{3}{4}(\partial g+ 2C \partial f)^2 g_{mn} \Big) \nn \\
 +4 e^{-12U-\phi}\Big(\partial_m f \partial_n f -\frac{3}{4}(\partial f)^2 g_{mn}\Big)\Big],\\
\square U+6 (\partial U)^2 + 5 e^{-2U} -\frac{1}{8}e^{-12U-\phi}(\partial f)^2-\frac{1}{32}e^{-12U+\phi}(\partial g+ 2C \partial f)^2=0, \label{U-eom} \\
\square \phi + 6\, \partial U \cdot \partial \phi-e^{2\phi}(\partial C)^2-\frac{1}{2}e^{-12U-\phi}(\partial f)^2+\frac{1}{8}e^{-12U+\phi}(\partial g+ 2C \partial f)^2=0,\label{phi-eom}\\
\square C+ 6\, \partial U\cdot \partial C+ 2\, \partial \phi\cdot \partial C
+\frac{1}{2}e^{-12U-\phi}(\partial f)\cdot(\partial g+ 2C \partial f)=0\label{C-eom}, \\
\partial \Big(\sqrt{g_4}e^{-6U-\phi}\Big(\partial f +\frac{1}{2}e^{2\phi}C(\partial g+2C \partial f)\Big)\Big)=0,\label{f-eom}\\
\partial\Big(\sqrt{g_4}e^{-6U+\phi}(\partial g+ 2C \partial f)\Big)=0\label{g-eom}.
\end{align}
One can study the integrability conditions of the Killing spinor equations and check whether the supersymmetry conditions satisfy the equations of motion and the Bianchi identities automatically. Instead, here we checked that the metric \eqref{metric0} and the solutions to the BPS equations \eqref{BPS3-axion}--\eqref{BPS5}, do satisfy the above equations of motion.

\vspace{1cm}

\section{Four-dimensional effective action} \label{nlsm}

\subsection{Non-Linear Sigma Model}

In this section we study AdS$_6$ solutions of type IIB supergravity from a different perspective $i.e.$  by performing a dimensional reduction of type IIB supergravity on AdS$_6$ space to a four-dimensional theory. From the equations of motion obtained in the previous section, we construct a four-dimensional effective Lagrangian as
\begin{align}
{\cal{L}}=\sqrt{g_4} \,e^{6U}
\Big[ R&+30 (\partial U)^2
-\frac{1}{2}(\partial \phi)^2-\frac{1}{2}e^{2\phi}(\partial C)^2\\
&+\frac{1}{2}e^{-12U-\phi}(\partial f)^2+\frac{1}{8}e^{-12U+\phi}(\partial g+ 2C\partial f)^2 -30e^{-2U}\Big].\nn
\end{align}
By rescaling the metric $g_{mn}= e^{-6U}\tilde{g}_{mn}$, we have the Einstein frame Lagrangian
\begin{align}
{\cal{L}}=\sqrt{\tilde{g_4}}  \Big[ \tilde{R}&-24 (\partial U)^2
-\frac{1}{2}(\partial \phi)^2-\frac{1}{2}e^{2\phi}(\partial C)^2\nn \\
&+\frac{1}{2}e^{-12U-\phi}(\partial f)^2+\frac{1}{8}e^{-12U+\phi}(\partial g+ 2C\partial f)^2 -30e^{-8U}
\label{lag}
\Big],\\
=\sqrt{\tilde{g_4}}  \Big[ \tilde{R}&-\frac{1}{2} G_{IJ}\partial \Phi^I \partial \Phi^J-
V(\Phi)\Big],
\end{align}
where $\Phi^I, I=1,\cdots,5,$ are the five scalar fields $U,\, \phi,\, C,\, f$ and $g$. This is a non-linear sigma model of five scalar fields coupled to gravity with a non-trivial scalar potential. Note that the sign of the kinetic terms of the dualized scalars $f$ and $g$ is reversed. However it is well known that when we perform dimensional reduction on an internal space including time, the sign of certain kinetic terms come out reversed, $e.g.$ \cite{Cremmer:1998em}.

\subsection{Scalar kinetic terms}

We study properties of the five-dimensional target space. The metric is given by
\be
ds_5^2 = 48 dU^2 +d\phi^2 +e^{2\phi}dC^2-\frac{1}{4}e^{-12U+\phi}(dg+2Cdf)^2
-e^{-12U-\phi}df^2.\label{nlsm-metric}
\ee
This space is Einstein, which satisfies $R_{IJ}=-\frac{3}{2}G_{IJ}$. 

The dilaton $\phi$ and the axion $C$ form a complex one-form $P$. Also $g$ and $f$ originate from the complex three-form flux $G$. Hence, we turn to the four-dimensional sub-manifold spanned by $\phi,\, C,\, g,\, f$. We choose the orthonormal frame as
\begin{alignat}{2}
e^1&= d\phi ,& \qquad e^2&=e^\phi dC, \nn \\
e^3&= \frac{1}{2}e^{-6U+\phi/2}(dg+2Cdf),&\qquad  e^4&=e^{-6U-\phi/2}df,
\end{alignat}
and construct a $(1,1)$-form $J$ and a $(2,0)$-form $\Omega$
\bea
J&=&e^1 \wedge e^2+ e^3\wedge e^4, \nn\\
\Omega&= &(e^1+i e^2) \wedge (e^3+i e^4),
\eea
which satisfy
\be
J \wedge J= \frac{1}{2}\Omega \wedge \bar{\Omega}, \qquad J \wedge \Omega=0.
\ee
By taking an exterior derivative to these two-forms, we have
\bea
 dJ &=& 0, \\
d\Omega  &=& i P \wedge \Omega,
\eea
where $P=-\frac{3}{2}e^2$. Hence, we find that the four-dimensional submanifold is K\"ahler. Its Ricci form is obtained by ${\cal{R}}=dP=-\frac{3}{2}e^1\wedge e^2$.

To investigate the isometry of the target space, we solved the Killing equation $\nabla_{(I} K_{J)}=0$, and found eight Killing vectors in \eqref{8KV}. These Killing vectors generate an $sl(3,\mathbb{R})$ algebra. The details can be found in appendix \ref{KV-SL3}.

One can explicitly check that the five-dimensional target space is in fact the coset $\textrm{SL}(3,\mathbb{R})/\textrm{SO}(2,1)$.\footnote{Having a coset after dimensional reduction is of course a very familiar story in supergravity. As it is very well known, Kaluza-Klein reduction of $D=4$ Einstein gravity on a circle leads to $\textrm{SL}(2,\mathbb{R})/\textrm{SO}(2)$, and its bigger versions appear in various supergravity theories \cite{Cremmer:1998em, Maison:1979kx, Giusto:2007fx,  Fre:2009et}.} We construct the coset representative $\cal{V}$ in Borel gauge by exponentiating Cartan generators $H_1, H_2$ and positive root generators $E_{\alpha_1}, E_{\alpha_2}, E_{\alpha_3}$, 
\be
{\cal{V}}=e^{\frac{1}{\sqrt{2}}\phi H_1} e^{-2\sqrt{6}U H_2} e^{C E_{\alpha_1}} e^{f E_{\alpha_2}} 
e^{\frac{1}{2} g E_{\alpha_3}}.
\ee
With the basis of $\textrm{SL}(3,\mathbb{R})$ introduced in \eqref{SL(3R)mat}, one can obtain the coset representative ${\cal{V}}$ in a  $3 \times 3$ matrix form explicitly. Then we construct an element of the orthogonal complement of $so(2,1)$ in $sl(3,\mathbb{R})$,
\be
P_{\mu(ij)}= {\cal{V}}_{(i|}^{\phantom{(i}a} \partial_\mu({\cal{V}}^{-1})_a^{\phantom{a}k} \eta_{k|j)}.
\ee
Here $i,j,k=1,2,3$ is a vector index of $\textrm{SO}(2,1)$ and $a=1, 2, 3$ is an $\textrm{SL}(3,\mathbb{R})$ index. An invariant metric of $\textrm{SO}(2,1)$ is
\be
\eta_{ij}=\textrm{diag}(1,1,-1).
\ee
Finally, the kinetic terms of the scalar fields of the Lagrangian \eqref{lag} is
\be
{\cal{L}}_\textrm{kinetic}=-\textrm{Tr}(P_\mu P^\mu).
\ee

\subsection{Scalar potential}

Now let us consider the scalar potential $V=30 e^{-8U}$ in the Lagragian. Its existence must obviously break the $\textrm{SL}(3,\mathbb{R})$ global symmetry into a nontrivial subalgebra. Among the eight generators in appendix \ref{KV-SL3}, this scalar potential is invariant under the action of five Killing vectors $ K^1, K^3, K^4, K^6$ and $K^8$. With the following identification
\be
e_1= K^4, \quad e_2 = \sqrt{2}K^1,\quad e_3 =K^3,\quad e_4=K^6,\quad e_5=-K^8
\ee
one can see that they form a certain five-dimensional Lie algebra so-called $A_{5,40}$ in table II of \cite{Patera:1976ud}. This algebra is isomorphic to the semi-direct sum $sl(2,\mathbb{R})\ltimes \mathbb{R}^2$ \cite{Kato}.

The scalar potential here comes from the curvature of internal space AdS$_6$. Certainly the situation is very similar to gauged supergravities where the higher-dimensional origin of the gauging process is related to the curvature of the internal space. Within the context of lower-dimensional supergravity itself, compared to the un-gauged action, a subgroup of the global symmetry is made local and the associated vector fields acquire non-abelian gauge interactions. A new parameter, say $g$, should be introduced as gauge coupling. To preserve supersymmetry, the action and the supersymmetry transformations are modified and importantly for us in general a scalar potential should be added at order $g^2$. Although our theory is not a supergravity theory {\it per se}, and there are no vector fields, we borrow the idea of gauged supergravity and write the scalar potential in terms of the coset representative of non-linear sigma model, through the so-called $T$-tensor. This may be justified because our four-dimensional action also has Killing spinor equations which are compatible with the field equations. In other words the integrability condition of Killing spinor equations should imply the fields satisfy the Euler-Lagrange equations. It is the $T$-tensor which encodes the gauging process and determines the modification of supersymmetry transformation rules and the action in gauged supergravity.

For a class of the maximal supergravity theories with a global symmetry group $\textrm{SL}(n,\mathbb{R})$, it is well known that the gauged supergravity can be obtained by gauging the $\textrm{SO}(n)$ subgroup. This gauging can be generalized to the non-compact subgroup $\textrm{SO}(p,q)$ with $p+q=n$ and the non-semi-simple group $\textrm{CSO}(p,q,r)$ with $p+q+r=n$, which was introduced in \cite{Hull:1984vg, Hull:1984qz}. $\textrm{CSO}(p,q,r)=\textrm{SO}(p,q)\ltimes \mathbb{R}^{(p+q)\cdot r}$ is a subgroup of $\textrm{SL}(n,\mathbb{R})$, {e.g.} (6.8) of \cite{Samtleben:2005bp}, and preserve the metric 
\be
q_{ab}= \textrm{diag}(\underbrace{1, \cdots}_p, \underbrace{-1, \cdots}_q, \underbrace{0, \cdots}_r).
\ee

Let us focus on the non-semi-simple group CSO$(1,1,1)$. We introduce the $T$-tensor as (apparently in the same way as in the gauged supergravity)
\be
T_{ij}={\cal{V}}_i^{\phantom{i}a}{\cal{V}}_j^{\phantom{i}b} q_{ab},
\ee
where
\be
q_{ab}= \textrm{diag} (1,-1,0).
\ee
Then one can easily check that the scalar potential is
\be
V=-15\Big((\textrm{Tr}T)^2-\textrm{Tr}(T^2)\Big).
\ee
It should be possible to re-write the Killing spinor equations \eqref{dif-gra1}--\eqref{alg-dil2} as well as the action to make the symmetry $\textrm{SL}(3,\mathbb{R})$ and the choice of CSO$(1,1,1)$ more manifest. We plan to do this construction, based on Killing spinor equations and their compatibility with the field equations, for all possible choices of compact and non-compact maximal subgroups of $\textrm{SL}(n,\mathbb{R})$ in a separate publication.

\section{Discussions}

We have studied AdS$_6$ solutions of type IIB supergravity theory in this paper.
In the first part, we have employed the Killing spinor analysis and revisited supersymmetric AdS$_6$ solutions, which was studied in \cite{Apruzzi:2014qva} using the pure spinor approach. We have constructed three Killing vectors, which satisfy $SU(2)$ algebra and give $S^2$ factor in the four-dimensional internal space $M_4$.  In other words, the $SU(2)$ symmetry, which corresponds to $SU(2)_R$ R-symmetry in the dual field theory, appears as isometries of the background if we impose the supersymmetric conditions. Also we have found two one-forms which are orthogonal to the Killing vectors. Using these one-forms, we have introduced the coordinates and determined the metric of the remaining two-dimensional space, and two coupled PDEs defined on it. Also the scalar fields and three-form fluxes have been found. Once we are given the solution to the PDEs, then the metric and the fluxes can be determined. Our results completely agree with the work of \cite{Apruzzi:2014qva}. 

Although the result of \cite{Apruzzi:2014qva} makes a significant progress in the classification of the supersymmetric AdS$_6$ solutions in type IIB supergravity theory, there still remain a couple of important problems to be studied further. 
To be sure, the most important but difficult task is to solve the PDEs \eqref{BPS1}, \eqref{BPS2} and find a new AdS$_6$ solution. Also it is very important to construct the field theories dual to AdS$_6$ solutions of IIB supergravity, which is still unknown. In \cite{Lozano:2013oma}, the properties of the dual field theory were studied through their AdS$_6$ solution. For the general class of solutions studied in \cite{Apruzzi:2014qva}, the authors suggested that $(p,q)$ five-brane webs \cite{Aharony:1997ju} play a crucial role. They conjectured that $(p,q)$ five-brane webs might be somehow related to the PDEs and the supergravity solutions could be obtained in the near-horizon limit.

Our independent analysis adds credence to the fact that the nonlinear PDEs found in \cite{Apruzzi:2014qva} provide necessary and sufficient conditions for supersymmetric AdS$_6$ in IIB supergravity. One should however admit that the PDEs  in the present form are far from illuminating. As it is sometimes the case, the study of the general form of supersymmetric solutions in supergravity is not always very efficient in constructing new solutions. However, identifying the {\it canonical} form of the metric and form-fields as done in \cite{Apruzzi:2014qva} and in this paper are equivalent to having the complete information on Killing spinors. So they become very useful for the study of supersymmetric probe consideration, for instance in the study of supersymmetric Wilson loops from D-branes. 

We thus think that a less technical, and more intuitive way of understanding the supersymmetric AdS$_6$ solutions would be very desirable. We hope our analysis in the second half of this paper is a modest first step towards such framework. There
we have presented a four-dimensional theory via a dimensional reduction on AdS$_6$ space. The problem of finding AdS$_6$ solutions of type IIB supergravity is reduced to a four-dimensional non-linear sigma model, {\it i.e.} a gravity theory coupled to five scalars with a non-trivial scalar potential. The scalar kinetic terms parameterize $\textrm{SL}(3,\mathbb{R})/\textrm{SO}(2,1)$. And we have reconstructed the scalar potential in terms of the coset non-linear sigma model language in a manner inspired by the gauged supergravity. We discovered that a particular group CSO$(1,1,1)$ which is a subgroup of $\textrm{SL}(3,\mathbb{R})$ is relevant to the scalar potential at hand, and presented the analogue of $T$-tensor. We hope the knowledge of the symmetry structure in the effective four-dimensional action will become useful to get a deeper insight into the existing solutions \cite{Cvetic:2000cj,Lozano:2012au}, for the identification of
their gauge theory duals, and eventually also for constructing more explicit solutions.

The $D=4$ effective action at hand is purely bosonic and it is not expected to be part of a supergravity action. But it enjoys a nice property that it is equipped with an associated set of Killing spinor equations which allows BPS solutions. When the Killing spinor equations \eqref{dif-gra1}--\eqref{alg-dil2} are re-written in a covariant way where the coset symmetry and the choice of gauging group CSO$(1,1,1)$ is more manifest, we expect we can generalize the construction to a bigger symmetry $\textrm{SL}(n,\mathbb{R})$ with $n>3$ and also different choices of maximal subgroup thereof. Of course their string theory origin is not clear, but mathematically they are interesting ``fake supergravity" models and might be useful {\it e.g.} for bottom-up model building in the AdS/CFT inspired study of condensed matter physics. A similar generalization of BPS systems was successfully performed starting with AdS$_3$ solutions in IIB supergravity and AdS$_2$ solutions in eleven dimensional supergravity in the line of works reported in 
\cite{Kim:2005ez,Kim:2006qu,Gauntlett:2007ts}. We plan to report on such generic analysis in a separate publication.

\bigskip
\leftline{\bf Acknowledgements}
We are grateful to Dario Rosa for explaining the work \cite{Apruzzi:2014qva} to us. We thank Hiroaki Nakajima and Hoil Kim for comments and discussions. MS thanks Changhyun Ahn, Kimyeong Lee, and Jeong-Hyuck Park for encouragement and support. The work of MS was mostly done when he was a post-doctoral fellow at Korea Institute for Advanced Study. This research was supported by a post-doctoral fellowship grant from Kyung Hee University (KHU-20131358, HK and NK), National Research Foundation of Korea (NRF) grants funded by the Korea government (MEST) with grant No. 2010-0023121 (HK, NK, MS), No. 2012046278 (NK), No. 2013064824 (HK), 2013R1A1A1A05005747 (MS), and No. 2012-045385/2013- 056327/2014-051185 (MS).

\appendix

\section{Type IIB supergravity}

We follow the conventions of \cite{Gauntlett:2005ww}. In type IIB supergravity, the bosonic fields are the graviton $g_{MN}$, five-form flux $F_{(5)}$, complex three-form flux $G_{(3)}$, dilaton $\phi$ and axion $C$. For the fermionic fields, there are gravitino $\psi_M$ and dilatino $\lambda$. The supersymmetry variation of the fermionic fields are given by
\begin{align}
\delta\,\psi_M\,=&\,D_M\,\epsilon\,+\,\frac{1}{96}\,(\Gamma_M \,\Gamma\,^{NPQ}\,G_{NPQ}\,+\,2\,\Gamma^{NPQ}\,G_{NPQ}\,\Gamma_M)\,\epsilon^c\,  \\
& \quad \quad+\,\frac{i}{1920}\,\Gamma^{NPQRS}\,F_{NPQRS}\,\Gamma_M\,\epsilon\,, \nn \\
\delta\,\lambda\,=&\,i\,\Gamma^M\,P_M\,\epsilon^c\,+\,\frac{i}{24}\,\Gamma^{MNP}\,G_{MNP}\,\epsilon\,.
\end{align}
where the covariant derivative is 
\be
D_M \epsilon=(\nabla_M - \dfrac{i}{2} Q_M)\epsilon.
\ee
The fields $P_M$ and $Q_M$ are written in terms of the dilaton and axion as
\begin{align}
P &= \dfrac{i}{2} e^\phi d C +\dfrac{1}{2} d \phi , \nn \\
Q&= -\dfrac{1}{2} e^\phi d C.\label{PQ}
\end{align}
The chirality conditions are
\be
\Gamma_{11}\,\psi\,=\,-\,\psi\,, \qquad \Gamma_{11}\,\lambda\,=\,\lambda\,, \qquad \Gamma_{11}\,\epsilon\,=\,-\,\epsilon\,.
\ee

\section{Gamma matrices and spinors}

\subsection{Gamma matrices}

We follow the conventions of \cite{Sohnius:1985qm}. We decompose the ten-dimensional gamma matrices by writing
\begin{align} \label{gamma}
\Gamma_\mu\,&=\,\rho_\mu \otimes\gamma_5\,, \notag \\
\Gamma_m\,&=\,\,1\,\,\otimes\gamma_m\,,
\end{align}
where $\mu\,=\,0,1,2,3,4,5$ and $m\,=\,1,2,3,4$. Then the chirality matrix is given by $\Gamma_{11}\,=\,\rho_7\,\otimes\,\gamma_5$.

In even dimensions, we introduce the intertwiners, which act on the gamma matrices as
\begin{align} \label{int1}
A\,\Gamma_M\,A^{-1}\,=&\,\Gamma_M^\dagger\,, \notag \\
C^{-1}\,\Gamma_M\,C\,=&\,-\,\Gamma_M^T\,, \notag \\
D^{-1}\,\Gamma_M\,D\,=&\,-\,\Gamma_M^*\,, 
\end{align}
with $D\,=\,C\,A^T$. These intertwiners can be chosen to satisfy the following relations at given $d$ dimensions,
\begin{equation}
 A_d\,=\,A_d^\dagger\,, \qquad C_d\,=\,\eta\,C_d^T\,, \qquad D_d\,=\,\delta\,(D_d^*)^{-1},
\end{equation}
where the values of $\eta$ and $\delta$ are given in table \ref{eta}. We decompose the ten-dimensional intertwiners as
\bea
A_{10}\,=\,A_6\otimes A_4\,, \qquad C_{10}\,=\,C_6\otimes C_4\,, \qquad D_{10}\,=\,D_6\otimes D_4.
\eea

\begin{table}
\centering
\begin{tabular}{c |c c c}
$d$         & 4      & 6      & 10 \\
\hline
$\eta $    &  $-$ & $+$  &  $-$ \\
$\delta  $ &  $-$ & $-$   & $+$ \\
\end{tabular}
\caption{The values of $\eta$ and $\delta$ in various dimensions.}
\label{eta}
\end{table}
 
\subsection{Spinors}

There are two ten-dimensional Majorana-Weyl spinor $\epsilon_i$, which satisfy
\be
\Gamma_{11} \epsilon_i=-\epsilon_i, \qquad \epsilon_i^c=\epsilon_i,
\ee
where $i=1, 2$.
We decompose $\epsilon_i$ into six- and four-dimensional spinors, $\psi$ and $\chi$, respectively, as
\be
\epsilon_i = \psi_+ \otimes \chi_{i-} +\psi_-\otimes \chi_{i+}+ c.c. ,
\ee
where $\pm$ represent the chirality. In our case, the six-dimensional spinors $\psi_\pm$ satisfy the Killing spinor equation on $AdS_6$
\be
\nabla_\mu\,\psi_\pm\,=\,\frac{i}{2}\,m\,\rho_\mu\,\rho_7\,\psi_\mp,
\ee
where $m$ is the inverse radius of $AdS_6$. Then we have the complexified spinor,
\begin{align}\label{spinor}
\epsilon &\equiv \epsilon_1 + i  \epsilon_2, \nn \\ 
              &    =    \psi_+ \otimes \xi_{1-} +\psi_-\otimes \xi_{1+}+ \psi_+^c \otimes \xi_{2-}^c +\psi_-^c\otimes \xi_{2+}^c, 
\end{align}
where 
\be
\xi_{1\pm}=\chi_{1\pm}+i \chi_{2\pm},\qquad \xi_{2\pm}^c=\chi_{1\pm}^c+i \chi_{2\pm}^c.
\ee
The Dirac adjoint and the charge conjugation are, respectively 
\be
\bar{\eta}= \eta^\dagger A, \qquad \eta^c= D \eta^*.
\ee

\section{Spinor bilinears}

One can construct all the spinor bilinears such as $\overline{\xi}_{A, i} \gamma^{(a)} \xi_{B, j}$ and $ \overline{\xi^c_A}_{, i} \gamma^{(a)} \xi_{B, j}$. Here $A, B=1,2$ and $i, j$ represent the chirality $+, -$ and 
$\gamma^{(a)} \equiv\gamma^{m_1 \cdots m_a}$. Some of the spinor bilinears identically vanish by the chirality,
\be
\overline{\xi}_+ \xi_- =0, \qquad \overline{\xi}_+ \gamma_{m}\xi_+ =0, \qquad \overline{\xi}_+ \gamma_{mn}\xi_-=0.
\ee
Also due to the antisymmetry of the charge conjugation matrix $C_4$, we have
\begin{gather}
\overline{\xi^c_{1}}_+ \xi_{1+}=\overline{\xi^c_{2}}_+ \xi_{2+}=0, \qquad
\overline{\xi^c_{1}}_+ \xi_{2+}=-\overline{\xi^c_{2}}_+ \xi_{1+}, \qquad 
\overline{\xi^c_{1}}_+ \gamma_m \xi_{2-}=\overline{\xi^c_{2}}_- \gamma_m \xi_{1+}.
\end{gather}

\subsection{Algebraic relations} \label{alg-rel}

In this section, we study the algebraic relations between the spinor bilinears, which can be derived from the algebraic Killing equations \eqref{alg-gra1}--\eqref{alg-dil2}.

If we multiply $\overline{\xi}_{1\mp}$ to \eqref{alg-gra1} and $\xi_{2\mp}$ to a hermitian conjugate of \eqref{alg-gra2} , then eliminate the three-form flux terms, we have
\begin{gather}
\overline{\xi}_{1+}\xi_{1+}-\overline{\xi}_{2-}\xi_{2-} =-\overline{\xi}_{1-}\xi_{1-}+\overline{\xi}_{2+}\xi_{2+},\\
\partial_m U \Big(\overline{\xi}_{1+}\gamma^{n}\xi_{1-}+\overline{\xi}_{2+}\gamma^{n}\xi_{2-}+
\overline{\xi}_{1-}\gamma^{m}\xi_{1+}+\overline{\xi}_{2-}\gamma^{m}\xi_{2+}\Big)=0\label{du.vec}.
\end{gather}
If we multiply the charge conjugate spinor instead and follow the same procedure, we obtain
\begin{gather}
\overline{\xi^c_{2}}_+ \xi_{1+} = \overline{\xi^c_1}_- \xi_{2-},\label{SC12R}\\
\partial_m U \Big(\overline{\xi^c_1}_{+}\gamma^{m}\xi_{2-}+\overline{\xi^c_2}_{+}\gamma^{m}\xi_{1-} \Big)=0\label{du.cvec}.
\end{gather}
Similarly eliminating the terms which have only one gamma matrix, we obtain
\begin{gather}
\overline{\xi}_{2+} \xi_{1+}+\overline{\xi}_{2-} \xi_{1-}=0,\label{S12R}\qquad
\overline{\xi}_{1+} \xi_{2+}+\overline{\xi}_{1-} \xi_{2-}=0, \\
G_{mnp}\Big(\overline{\xi}_{1+} \gamma^{mnp}\xi_{1-}+\overline{\xi}_{2+} \gamma^{mnp}\xi_{2-}\Big)=0.
\end{gather}

\subsection{Differential relations} \label{diff-rel}

\leftline{\bf Scalar bilinears}
\begin{alignat} {2}
&\nabla_m(\overline{\xi}_{1\pm}\xi_{1\pm} &&+\overline{\xi}_{2\pm}\xi_{2\pm})=\partial_mU(\overline{\xi}_{1\pm}\xi_{1\pm}+\overline{\xi}_{2\pm}\xi_{2\pm})\label{S+}\\
& && +\frac{1}{2}ime^{-U}(\overline{\xi}_{1\pm}\gamma_m\xi_{1\mp}-\overline{\xi}_{1\mp}\gamma_m\xi_{1\pm}+\overline{\xi}_{2\pm}\gamma_m\xi_{2\mp}-\overline{\xi}_{2\mp}\gamma_m\xi_{2\pm})\,, \nn \\
&\nabla_m(\overline{\xi}_{1\pm}\xi_{1\pm} &&-\overline{\xi}_{2\pm}\xi_{2\pm})=-3\partial_mU(\overline{\xi}_{1\pm}\xi_{1\pm}-\overline{\xi}_{2\pm}\xi_{2\pm})\label{S-}\\
& &&-\frac{3}{2}ime^{-U}(\overline{\xi}_{1\pm}\gamma_m\xi_{1\mp}-\overline{\xi}_{1\mp}\gamma_m\xi_{1\pm}-\overline{\xi}_{2\pm}\gamma_m\xi_{2\mp}+\overline{\xi}_{2\mp}\gamma_m\xi_{2\pm})\,, \nn \\
&\nabla_m(\overline{\xi}_{2\pm}\xi_{1\pm}) &&=(i Q_m-\partial_mU)\overline{\xi}_{2\pm}\xi_{1\pm}-P_m\overline{\xi}_{1\pm}\xi_{2\pm}\label{S21}\\
& &&-\frac{1}{2}ime^{-U}(\overline{\xi}_{2\pm}\gamma_m\xi_{1\mp}-\overline{\xi}_{2\mp}\gamma_m\xi_{1\pm})\,,\nn \\
&\nabla_m(\overline{\xi_2^c}_\pm\xi_{1\pm}) &&=-\partial_mU\overline{\xi_2^c}_\pm\xi_{1\pm}+2\partial_mU\overline{\xi^c_1}_{\pm}\xi_{2\pm}\\
& &&-\frac{1}{2}ime^{-U}(\overline{\xi^c_2}_{\pm}\gamma_m\xi_{1\mp}-\overline{\xi^c_2}_{\mp}\gamma_m\xi_{1\pm})
+ime^{-U}(\overline{\xi^c_1}_{\pm}\gamma_m\xi_{2\mp}-\overline{\xi^c_1}_{\mp}\gamma_m\xi_{2\pm})\,. \nn
\end{alignat}
\leftline{\bf Vector biliears}
\begin{alignat}{2}
&\nabla^{[l}(\overline{\xi}_{1+}\gamma^{m]}\xi_{1-} &&+\overline{\xi}_{2+}\gamma^{m]}\xi_{2-})
=-6\,\partial^{[l}U(\overline{\xi}_{1+}\gamma^{m]}\xi_{1-}+\overline{\xi}_{2+}\gamma^{m]}\xi_{2-})\label{dK3} \\
& &&+\dfrac{3}{2}ime^{-U}(\overline{\xi}_{1+}\gamma^{lm}\xi_{1+}+\overline{\xi}_{1-}\gamma^{lm}\xi_{1-}+\overline{\xi}_{2+}\gamma^{lm}\xi_{2+}+\overline{\xi}_{2-}\gamma^{lm}\xi_{2-}), \nn \\  
&\nabla^{[l}(\overline{\xi}_{1+}\gamma^{m]}\xi_{1-} &&-\overline{\xi}_{2+}\gamma^{m]}\xi_{2-})
=-2\,\partial^{[l}U(\overline{\xi}_{1+}\gamma^{m]}\xi_{1-}-\overline{\xi}_{2+}\gamma^{m]}\xi_{2-}) \\
& &&+\dfrac{1}{2}ime^{-U}(\overline{\xi}_{1+}\gamma^{lm}\xi_{1+}+\overline{\xi}_{1-}\gamma^{lm}\xi_{1-}-\overline{\xi}_{2+}\gamma^{lm}\xi_{2+}-\overline{\xi}_{2-}\gamma^{lm}\xi_{2-}),\nn \\
&\nabla^{[l}(\overline{\xi}_{2+}\gamma^{m]}\xi_{1-})
&&=(i Q^{[l}-4\,\partial^{[l}U)\overline{\xi}_{2+}\gamma^{m]}\xi_{1-} 
+P^{[l}(\overline{\xi}_{1+}\gamma^{m]}\xi_{2-}) \label{d(iL2-L3)}  \\
& &&+2ime^{-U}(\overline{\xi}_{2+}\gamma^{lm}\xi_{1+}+\overline{\xi}_{2-}\gamma^{lm}\xi_{1-}),\nn\\
&\nabla^{[l}(\overline{\xi}_{1+}\gamma^{m]}\xi_{2-})
&&=(-i Q^{[l}-4\,\partial^{[l}U)\overline{\xi}_{1+}\gamma^{m]}\xi_{2-}+P^{*[l}(\overline{\xi}_{2+}\gamma^{m]}\xi_{1-}) \label{d(iL2+L3)} \\
& &&+2ime^{-U}(\overline{\xi}_{1+}\gamma^{lm}\xi_{2+}+\overline{\xi}_{1-}\gamma^{lm}\xi_{2-}),\nn\\
&\nabla^{[l}(\overline{\xi_1^c}_+\gamma^{m]}\xi_{1-})
&&=(i Q^{[l}-4\partial^{[l}U) \overline{\xi_1^c}_+\gamma^{m]}\xi_{1-}
+P^{[l}\overline{\xi_2^c}_+\gamma^{m]}\xi_{2-} \label{dVC11} \\
& &&+ime^{-U}(\overline{\xi^c_1}_+\gamma^{lm}\xi_{1+}+\overline{\xi^c_1}_-\gamma^{lm}\xi_{1-})\,, \nn\\ 
&\nabla^{[l}(\overline{\xi_2^c}_+\gamma^{m]}\xi_{2-})
&&=(-i Q^{[l}-4\partial^{[l}U) \overline{\xi_2^c}_+\gamma^{m]}\xi_{2-}+P^{*[l}\overline{\xi_1^c}_+\gamma^{m]}\xi_{1-} \label{dVC22}\\
& &&+ime^{-U}(\overline{\xi^c_2}_+\gamma^{lm}\xi_{2+}+\overline{\xi^c_2}_-\gamma^{lm}\xi_{2-})\,,\nn \\
&\nabla^{[l}(\overline{\xi_1^c}_+\gamma^{m]}\xi_{2-}&&+\overline{\xi_2^c}_+\gamma^{m]}\xi_{1-})
=-6\,\partial^{[l}U(\overline{\xi_1^c}_+\gamma^{m]}\xi_{2-}+\overline{\xi_2^c}_+\gamma^{m]}\xi_{1-} )
\label{d(K1+iK2)} \\
& &&+\dfrac{3}{2}ime^{-U}(\overline{\xi^c_1}_{+}\gamma^{lm}\xi_{2+}+\overline{\xi^c_1}_-\gamma^{lm}\xi_{2-}+\overline{\xi^c_2}_{+}\gamma^{lm}\xi_{1+}+\overline{\xi^c_2}_-\gamma^{lm}\xi_{1-})\,, \nn\\ 
&\nabla^{[l}(\overline{\xi_1^c}_+\gamma^{m]}\xi_{2-}&&-\overline{\xi_2^c}_+\gamma^{m]}\xi_{1-})
=-2\,\partial^{[l}U(\overline{\xi_1^c}_+\gamma^{m]}\xi_{2-}-\overline{\xi_2^c}_+\gamma^{m]}\xi_{1-} ) \\
& && +\dfrac{1}{2}ime^{-U}(\overline{\xi^c_1}_{+}\gamma^{lm}\xi_{2+}+\overline{\xi^c_1}_-\gamma^{lm}\xi_{2-}-\overline{\xi^c_2}_{+}\gamma^{lm}\xi_{1+}-\overline{\xi^c_2}_-\gamma^{lm}\xi_{1-})\,.\nn
\end{alignat}
\leftline{\bf Two-form bilinears}
\begin{alignat}{2}
&\nabla^{[r}(\overline{\xi}_{1\pm}\gamma^{st]}\xi_{1\pm}&&+\overline{\xi}_{2\pm}\gamma^{st]}\xi_{2\pm})
=-5\partial^{[r}U(\overline{\xi}_{1\pm}\gamma^{st]}\xi_{1\pm}+\overline{\xi}_{2\pm}\gamma^{st]}\xi_{2\pm})  \\
& &&-\frac{5}{6}ime^{-U}(\overline{\xi}_{1\pm}\gamma^{rst}\xi_{1\mp}-\overline{\xi}_{1\mp}\gamma^{rst}\xi_{1\pm}+\overline{\xi}_{2\pm}\gamma^{rst}\xi_{2\mp}-\overline{\xi}_{2\mp}\gamma^{rst}\xi_{2\pm})\,,\nn\\
&\nabla^{[r}(\overline{\xi}_{1\pm}\gamma^{st]}\xi_{1\pm}&&-\overline{\xi}_{2\pm}\gamma^{st]}\xi_{2\pm})
=-\partial^{[r}U(\overline{\xi}_{1\pm}\gamma^{st]}\xi_{1\pm}-\overline{\xi}_{2\pm}\gamma^{st]}\xi_{2\pm})\\
& &&+\frac{1}{3}(G^{rst}\overline{\xi}_{1\pm}\xi_{2\pm}-G^{*rst}\overline{\xi}_{2\pm}\xi_{1\pm}) \notag \\
& &&-\frac{1}{6}ime^{-U}(\overline{\xi}_{1\pm}\gamma^{rst}\xi_{1\mp}-\overline{\xi}_{1\mp}\gamma^{rst}\xi_{1\pm}-\overline{\xi}_{2\pm}\gamma^{rst}\xi_{2\mp}+\overline{\xi}_{2\mp}\gamma^{rst}\xi_{2\pm})\,,\nn \\
&\nabla^{[r}(\overline{\xi}_{2\pm}\gamma^{st]}\xi_{1\pm})
&&=(i Q^{[r}-3\partial^{[r}U)\overline{\xi}_{2\pm}\gamma^{st]}\xi_{1\pm}+P^{[r}\overline{\xi}_{1\pm}\gamma^{st]}\xi_{2\pm}  \\
& &&-\frac{1}{6}G^{rst}(\overline{\xi}_{1\pm}\xi_{1\pm}-\overline{\xi}_{2\pm}\xi_{2\pm})-\frac{1}{2}ime^{-U}(\overline{\xi}_{2\pm}\gamma^{rst}\xi_{1\mp}-\overline{\xi}_{2\mp}\gamma^{rst}\xi_{1\pm})\,,\nn \\
&\nabla^{[r}(\overline{\xi_1^c}_\pm\gamma^{st]}\xi_{1\pm})
&&=(i Q^{[r}-3\partial^{[r}U )\overline{\xi_1^c}_\pm\gamma^{st]}\xi_{1\pm}+P^{[r}\overline{\xi_2^c}_\pm\gamma^{st]}\xi_{2\pm}+\frac{1}{6}G^{rst}(\overline{\xi_1^c}_\pm\xi_{2\pm}-\overline{\xi_2^c}_\pm\xi_{1\pm}) \notag \label{dTC11}\\
& &&-\frac{1}{2}ime^{-U}(\overline{\xi^c_1}_{\pm}\gamma^{rst}\xi_{1\mp}-\overline{\xi^c_1}_{\mp}\gamma^{rst}\xi_{1\pm})\,,  \\
&\nabla^{[r}(\overline{\xi_2^c}_\pm\gamma^{st]}\xi_{2\pm})
&&=(-i Q^{[r}-3\partial^{[r}U)\overline{\xi_2^c}_\pm\gamma^{st]}\xi_{2\pm}+P^{*[r}\overline{\xi_1^c}_\pm\gamma^{st]}\xi_{1\pm}-\frac{1}{6}G^{*rst}(\overline{\xi_1^c}_\pm\xi_{2\pm}-\overline{\xi_2^c}_\pm\xi_{1\pm}) \notag \label{dTC22}\\
& &&-\frac{1}{2}ime^{-U}(\overline{\xi^c_2}_{\pm}\gamma^{rst}\xi_{2\mp}-\overline{\xi^c_2}_{\mp}\gamma^{rst}\xi_{2\pm})\,, \\
&\nabla^{[r}(\overline{\xi_2^c}_\pm\gamma^{st]}\xi_{1\pm})
&&=-3\partial^{[r}U\overline{\xi_2^c}_\pm\gamma^{st]}\xi_{1\pm}-\partial^{[r}(2U)\overline{\xi_1^c}_\pm\gamma^{st]}\xi_{2\pm}  \\
& &&-\frac{1}{2}ime^{-U}(\overline{\xi^c_2}_{\pm}\gamma^{rst}\xi_{1\mp}-\overline{\xi^c_2}_{\mp}\gamma^{rst}\xi_{1\pm})\,.\nn
\end{alignat}

\subsection*{Normalization of scalar bilinears} 

From \eqref{S+} and \eqref{S-}, we have
\be
d[e^{-U}(\overline{\xi}_{1+}\xi_{1+}+\overline{\xi}_{1-}\xi_{1-})]=
d[e^{-U}(\overline{\xi}_{2+}\xi_{2+}+\overline{\xi}_{2-}\xi_{2-})]=0.
\ee
Then, we can fix the normalization 
\be\label{sc-norm}
\overline{\xi}_{1+}\xi_{1+}+\overline{\xi}_{1-}\xi_{1-}=
\overline{\xi}_{2+}\xi_{2+}+\overline{\xi}_{2-}\xi_{2-}= \dfrac{e^U}{3m}.
\ee

\section{Fierz identities}

In four dimensions, the Fierz identity is
\begin{align}
\eta_1^T\,\eta_2\,\eta_3^T\,\eta_4\,=\,&\frac{1}{4}\,\left(\eta_1^T\,\eta_4\,\eta_3^T\,\eta_2\,+\eta_1^T\,\gamma_5\,\eta_4\,\eta_3^T\,\gamma_5\,\eta_2\right) \notag \\ 
+\,&\frac{1}{4}\,\left(\eta_1^T\,\gamma^m\,\eta_4\,\eta_3^T\,\gamma_m\,\eta_2\,-\,\eta_1^T\,\gamma^m\,\gamma_5\,\eta_4\,\eta_3^T\,\gamma_m\,\gamma_5\,\eta_2\,\right)\, \notag \\
-\,&\frac{1}{8}\,\eta_1^T\,\gamma^{mn}\,\eta_4\,\eta_3^T\,\gamma_{mn}\,\eta_2\,.
\end{align}
When we calculate the Lie bracket of the Killing vectors, we need to compute contractions of vectors with two-forms. With  the spinors $\eta_2$ and $\eta_3$ of the same chirality, we find the following relation useful 
\be\label{fierz1.2}
\eta_1^T\,\gamma^{m}\,\eta_4\,{\eta_3^T}\,\gamma_{mn}\,\eta_{2}\,
=2\, \eta_1^T\,\gamma_n\, \eta_2\,\eta_3^T\,\eta_4-2\, \eta_3^T\,\gamma_n\, \eta_4\,\eta_1^T\,\eta_2\,- \eta_1^T\,\gamma_n \,\gamma_5\, \eta_4\,\eta_3^T\,\gamma_5\,\eta_2.
\ee

\subsection{Relations of scalar bilinears}

We also find useful relations between the scalar bilinears using the Fierz identities. If we choose $\eta_1^T=\overline{\xi}_{1+},\, \eta_2=\xi_{2+},\,\eta_3^T=\overline{\xi}_{2+},\,\eta_4=\xi_{1+},$ we have
\be
\overline{\xi}_{1+} \xi_{2+}\, \overline{\xi}_{2+} \xi_{1+}
= \frac{1}{2} \overline{\xi}_{1+} \xi_{1+}\, \overline{\xi}_{2+} \xi_{2+}
-\frac{1}{8}\overline{\xi}_{1+} \gamma^{mn}\xi_{1+}\, \overline{\xi}_{2+}\gamma_{mn} \xi_{2+}.
\ee

Similarly, if we choose $\eta_1^T=\overline{\xi^c_1}_+,\, \eta_2=\xi_{2+},\,\eta_3^T=\overline{\xi}_{2+},\,\eta_4=\xi^c_{1+},$ we have
\be
\overline{\xi^c_1}_+\xi_{2+}\, \overline{\xi}_{2+} \xi^c_{1+}
= \frac{1}{2} \overline{\xi}_{1+} \xi_{1+}\, \overline{\xi}_{2+} \xi_{2+}
+\frac{1}{8}\overline{\xi}_{1+} \gamma^{mn}\xi_{1+}\, \overline{\xi}_{2+}\gamma_{mn} \xi_{2+}.
\ee
Thus we find that
\be\label{fierz-s}
|\overline{\xi}_{1+} \xi_{2+}|^2+|\overline{\xi^c_1}_+\xi_{2+}|^2=
\overline{\xi}_{1+} \xi_{1+}\, \overline{\xi}_{2+} \xi_{2+}.
\ee

We also have a similar result with the minus chirality spinors. Then, using \eqref{SC12R} and \eqref{S12R}, we obtain 
\be
\overline{\xi}_{1+} \xi_{1+}\, \overline{\xi}_{2+} \xi_{2+}=\overline{\xi}_{1-} \xi_{1-}\, \overline{\xi}_{2-}\xi_{2-}.
\ee 
With the normalization \eqref{sc-norm}, we conclude that 
\be
\overline{\xi}_{1+} \xi_{1+}=\overline{\xi}_{2-} \xi_{2-}, \qquad \overline{\xi}_{2+} \xi_{2+}=\overline{\xi}_{1-} \xi_{1-}.
\ee

\subsection{Inner products of vector bilinears} \label{innervec}

The vectors $K_1, K_2, K_3$ and one-forms $ L^2, L^3$ play a crucial role in determining the form of the four-dimensional metric. In this section we explain the procedure in detail. First, we calculate the norms and the inner products of these vectors using the Fierz identities. For example, choosing $\eta_1^T=\overline{\xi}_{1+},\, \eta_2=\xi_{2+},\ \eta_3^T=\overline{\xi}_{2-},\, \eta_4=\xi_{1-},$ we get
\be
\overline{\xi}_{1+}\gamma^{n}\,\xi_{1-}\,\overline{\xi}_{2-}\gamma_{n}\,\xi_{2+}=2\,\overline{\xi}_{1+}\xi_{2+}\,\overline{\xi}_{2-}\xi_{1-}.
\ee 
Similarly, inner products of any vectors can be written as products of scalars. For the three Killing vectors, we have
\bea\label{innerK}
(K_1)^2&=& (\overline{\xi}_{1+} \xi_{1+}-\overline{\xi}_{2+} \xi_{2+})^2-(\overline{\xi^c_1}_+\xi_{2+}-(\overline{\xi^c_1}_+\xi_{2+})^*)^2,\\
(K_2)^2&= &(\overline{\xi}_{1+} \xi_{1+}-\overline{\xi}_{2+} \xi_{2+})^2+(\overline{\xi^c_1}_+\xi_{2+}+(\overline{\xi^c_1}_+\xi_{2+})^*)^2,\nn\\
(K_3)^2&= &4|\overline{\xi^c_1}_+\xi_{2+}|^2,\nn\\
K_1 \cdot K_2 &=& i((\overline{\xi^c_1}_+\xi_{2+})^2-(\overline{\xi^c_1}_+\xi_{2+})^{*2}),\nn\\
K_1 \cdot K_3&=&(\overline{\xi}_{1+} \xi_{1+}-\overline{\xi}_{2+} \xi_{2+})(\overline{\xi^c_1}_+\xi_{2+}+(\overline{\xi^c_1}_+\xi_{2+})^*),\nn\\
K_2 \cdot K_3&=&-i (\overline{\xi}_{1+} \xi_{1+}-\overline{\xi}_{2+} \xi_{2+})(\overline{\xi^c_1}_+\xi_{2+}-(\overline{\xi^c_1}_+\xi_{2+})^*).\nn
\eea
The inner products of $L_2$ and $L_3$ are 
\begin{align}
(L^2)^2&=(\overline{\xi}_{1+}\xi_{1+}+\overline{\xi}_{2+}\xi_{2+})^2-(\overline{\xi}_{1+}\xi_{2+}+\overline{\xi}_{2+}\xi_{1+})^2 = \dfrac{1}{9m^2}e^{2U}(1-e^{-4U-\phi}y^2), \\
(L^3)^2&=(\overline{\xi}_{1+}\xi_{1+}+\overline{\xi}_{2+}\xi_{2+})^2+(\overline{\xi}_{1+}\xi_{2+}-\overline{\xi}_{2+}\xi_{1+})^2 = \dfrac{1}{9m^2}e^{2U}(1-e^{-4U+\phi}z^2), \\
L^2 \cdot L^3 &=-i\Big((\overline{\xi}_{1+}\xi_{2+})^2-(\overline{\xi}_{2+}\xi_{1+})^2\Big)=\frac{1}{9m^2}e^{-2U}\,y\,z,
\end{align}
where we express the scalar bilinears in terms of the coordinates $z$ and $y$ defined in \eqref{coord} at the last step.

\section{Killing vectors of five-dimensional target space} \label{KV-SL3}

We have found the eight Killing vectors of the five-dimensonal target space of the non-linear sigma model \eqref{nlsm-metric} as
\begin{align}\label{8KV}
K^1 &=\sqrt{2}(\partial_\phi -C\, \partial_C +\frac{f}{2}\, \partial_f -\frac{g}{2}\, \partial_g), \\
K^2 &=\frac{1}{2\sqrt{6}}( \partial_U +6f\,  \partial_f +6g\,  \partial_g), \nn \\
K^3 &=-\frac{1}{2}\,  \partial_C + f\,  \partial_g, \nn \\
K^4 &= -4C\,  \partial_\phi +2(C^2-e^{-2\phi})\, \partial_C +g\,  \partial_f,\nn\\
K^5 &= -\frac{1}{4}g\, \partial_U +(g+4 C f) \, \partial_\phi- e^{-2\phi}(-2f+2 C^2 e^{2\phi}f+C e^{2\phi}g)\,\partial_C \nn \\
       &+ (2Ce^{12U+\phi}-fg)\, \partial_f - e^{-\phi}(4e^{12U}+4C^2 e^{12U+2\phi}+e^{\phi}g^2\,\partial_g),\nn\\
K^6 &= \partial_g, \nn\\
K^7 &= -\frac{1}{4} f\, \partial_U -f\,  \partial_\phi +(\frac{1}{2}g+Cf)\, \partial_C-(e^{12U+\phi}+f^2)\, \partial_f-(fg-2C e^{12U+\phi}) \,\partial_g ,\nn\\
K^8 &= \partial_f.\nn
\end{align}

The eight generators of the $\textrm{SL}(3,\mathbb{R})$ group are
\be \label{SL(3R)mat}
T_1 \equiv H_1=\frac{1}{\sqrt{2}}\left( \begin{array}{ccc} 1& 0 & 0 \\0 & -1 & 0 \\0 & 0 & 0 
 \end{array}\right),~~
T_2\equiv H_2=\frac{1}{\sqrt{6}}\left( \begin{array}{ccc} 1 & 0 & 0 \\0 & 1 & 0 \\0 & 0 & -2 
 \end{array}\right),~~
\ee
\begin{equation}\nn
T_3 \equiv E_{\alpha_1}=\left( \begin{array}{ccc} 0 & 1 & 0 \\0 & 0 & 0 \\0 & 0 & 0 
 \end{array}\right),~~
T_5 \equiv E_{\alpha_2}=\left( \begin{array}{ccc} 0 & 0 & 0 \\0 & 0 & 1 \\0 & 0 & 0 
 \end{array}\right),~~
T_7 \equiv E_{\alpha_3}=\left( \begin{array}{ccc} 0 & 0 & 1\\0 & 0 & 0 \\0 & 0 & 0 
 \end{array}\right), \\
\ee
\be \nn
T_4 \equiv E_{-\alpha_1}=\left( \begin{array}{ccc} 0 & 0 & 0 \\1 & 0 & 0 \\0 & 0 & 0 
 \end{array}\right),~~
T_6 \equiv E_{-\alpha_2}=\left( \begin{array}{ccc} 0 & 0 & 0 \\0 & 0 & 0 \\0 & 1 & 0 
 \end{array}\right),~~
T_8 \equiv E_{-\alpha_3}=\left( \begin{array}{ccc} 0 & 0 & 0\\0 & 0 & 0 \\1& 0 & 0 
 \end{array}\right),
\ee
where $H_1, H_2$ are Cartan generators and $E_{\alpha_1}, E_{\alpha_2}, E_{\alpha_3}$ are positive root generators. By identifying $K^i=T_i$, the eight Killing vectors satisfy an $sl(3,\mathbb{R})$ algebra.



\end{document}